# The roles of adhesion, internal heat generation and elevated temperatures in normally loaded, sliding rough surfaces

Benjamin Poole [a+], Bartosz Barzdajn [a*], Daniele Dini [b], David Stewart [c], Fionn P. E. Dunne [a]

[a] Department of Materials, Imperial College, London, SW7 2AZ, UK

[b] Department of Mechanical Engineering, Imperial College, London, SW7 2AZ, UK

[c] Rolls-Royce plc., Raynesway, Derby, UK

[*] Now at School of Materials, University of Manchester, M1 3BB, UK

[+] Corresponding author, b.poole16@imperial.ac.uk

## Highlights

- Temperature rises due to plastic and heat generation are inconsequential when sliding rates representative to galling are accounted for

- The strength of adhesion is not relevant to the deformation of rough surfaces in the context of single-phase austenitic 316L stainless steel during metal-on-metal contact

- Increasing temperatures show quantifiable effects on galling resistance but not to the level as found experimentally in the literature

- Galling appears to be controlled by mechanisms (e.g. phase transformation, surface coating degradation) other than the simple mechanisms considered here

# Abstract

The thermal effects of plastic and frictional heat generation and elevated temperature were examined along with the role of adhesion in the context of galling wear, using a representative crystal plasticity, normally loaded, sliding surface model. Galling frequency behaviour was predicted for 316L steel. Deformation of the surfaces was dominated by the surface geometry, with no significant effect due to variations in frictional models. Plastic and frictional heating were found to have a minimal effect on the deformation of the surface, with the rapid conduction of heat preventing any highly localised heating. There was no corresponding effect on the predicted galling frequency response.

Isothermal, elevated temperature conditions caused a decrease in galling resistance, driven by the temperature sensitivity of the critical resolved shear stress. The extent of deformation, as quantified by the area of plastically deformed material and plastic reach, increased with temperature. Comparisons were made with literature results for several surface amplitude and wavelength conditions. Model results compared favourably with those in the literature. However, the reduction in predicted galling resistance with elevated temperature for a fixed surface was not as severe as observations in the literature, suggesting other mechanisms (e.g. phase transformations, surface coatings and oxides) are likely important.

**Keywords:** Sliding contact; Crystal plasticity; Finite element; Galling; Heat generation

# 1 Introduction

Galling is a severe plastic deformation process, occurring between highly loaded, sliding surfaces. The galling mechanism is associated with macroscopic plastic flow, surface roughening and the formation of protrusions [1]. This extensive surface roughening can lead to seizure of the surfaces, and accordingly galling is consequential to the operability of moving parts and machinery under load. Galling is typically a result of poor surface preparation, excessive load, high temperature and poor lubrication. Due to the elevated operating temperature and strict controls placed on coolant chemistry preventing the use of lubrication, a range of loaded surfaces susceptible to galling are found in pressurised water reactors (PWRs), typically within coolant pumps, control rod drive mechanisms and valves. Galling resistance is commonly achieved with the use of cobalt-based Stellite[1] hard facings [2].

The reduction of the use of cobalt-base alloys is an industry-wide goal for nuclear generation, due to both radiological issues associated with in-core activation of cobalt wear products and uncertainty in future cobalt prices [3,4]. During operation, cobalt wear products become entrained in the coolant flow, enter the reactor core and are activated by the high neutron fluxes. The resulting $Co^{60}$ (half-life 5.27 years) is a strong gamma-emitter and represents a significant contributor to the shutdown radiation field of the reactor [4]. Consequently, Co-free replacements are sought.

Work has proceeded for several decades to develop Co-free hard facings, focussing on Ni- and Fe-based alloys. Both galling and corrosion resistance are required at PWR operating temperatures (300°C). Attempts to develop alloys to meet the performance requirements have thus far not been completely successful with development alloys being unable to withstand loads sufficient to prevent the occurrence of galling or presenting manufacturing or corrosion issues [3,5–9]. Much of the difficulty has been in reproducing the elevated temperature performance of Stellite, with alternatives showing poor galling resistance at temperatures approaching 300°C [8,10–13]. Burdett [10] observed significant reductions in the load required to initiate galling in several stainless steel hard facings (Tristelle 5183, Delcrome 910 and 90, Nitronic 60). A related alloy, NOREM 02 (Electric Power Research Institute), has also been found to demonstrate significant temperature sensitivity, showing excellent performance at room temperature but a substantial decrease in galling resistance at higher temperatures [8,11,13,14]. Kim & Kim [8] attributed this behaviour to a change in wear mechanism, from low temperature oxidative wear to severe adhesive wear at 190°C and galling at higher temperatures. This change in wear mechanism was concluded to be due to a loss in work-hardening ability of the matrix through the high temperature suppression of the $\gamma \rightarrow \alpha'$ strain-induced martensitic transformation. Further work by Persson et al. [11] reported a similar degradation in galling resistance in NOREM 02 but could not confirm any loss in strain-induced martensitic transformation, reportedly due to the narrowness of the wear tracks resulting in weak x-ray diffraction signals.

---

[1] Stellite is a trade mark of Kennametal Inc.

This degradation in performance is not limited to hard facing steels. Austenitic stainless steels (e.g. AISI 3xx) show poor galling resistance in general, galling at loads below 10 MPa [15,16]. Whilst typically not suitable for industrial use, these steels could elucidate further details of the galling mechanism due to their similarities with various austenitic hard facings. Harsha et al. [17,18] reported low room temperature Galling$_{50}$ loads (the load at which half of sample would be expected to gall during ASTM G196 testing[2]) in this region for 304, 316, and their low carbon variants. Tests performed at 300°C showed significant reductions in Galling$_{50}$ for all four materials, attributed to a reduction in hardness.

Several mechanisms have been proposed to account for this temperature sensitivity, with several authors considering the roles of $\gamma \rightarrow \alpha'$ martensitic transformations [8,11,14,19,20], stacking fault energy [20–22], oxidation and oxide layer support [11,19] and the strength of adhesion [23,24].
Talonen and Hänninen [20] related the temperature sensitivity of α′-martensite formation to that of stacking fault energy, with the formation of stacking faults appearing to control the formation of α′-martensite. As such, these two mechanisms cannot be considered in isolation.

The formation of stacking faults results in work hardening, with several authors [22,25,26] suggesting that this type of hardening is key to galling resistance. Bhansali and Miller [22] proposed that a low stacking fault energy leads to galling resistance, through the action of stacking faults preventing cross-slip and enhancing work hardening of asperities. Whilst stacking fault formation is an important work hardening mechanism, the role of stacking faults in martensite formation may be of more importance to galling resistance than the hardening inherent in stacking fault formation itself.

Surface oxidation and contamination often provide low-friction, protective surface layers on components and will also demonstrate temperature sensitivity [10,22]. Oxide layers must be sufficiently adhered to and mechanically compatible with the substrate material in order to remain attached during sliding [27]. There are several examples in iron- and nickel-based alloys of oxide layers protecting surfaces provided there is sufficient support from the substrate [8,9,19,19,26,28]. Since oxidation and contamination will always be present in-service, the plastic deformation of surfaces at elevated temperatures will be critical in determining galling resistance.

---

[2] As in ASTM G196, galling frequency is defined as $F(l) = \frac{1}{1 + \exp\left(-\frac{l - G_{50}}{b}\right)}$ where $l$ is the applied load and $G_{50}$ is the Galling$_{50}$ parameter. It is clear that $F(\,l = G_{50}\,) = 0.5$. The factor $b$ is related to the shape of the distribution.

The strength of adhesion is typically observed to increase with temperature between bare affine metal surfaces. This has been investigated both theoretically (Rabinowicz [23]) and experimentally (Gåård et al. [24]). Adhesion is also known to be sensitive to crystal structure and orientation, with hcp material less susceptible to adhesion than bcc and fcc [24,29]. Buckley [29] proposed that the reduced potential for slip in hcp materials was the cause of this behaviour. Molinari et al. [30] demonstrated the influence of adhesion in the deformation of materials at the atomistic length scale, well below the length scale addressed by Barzdajn et al. [31], who investigated the deformation of engineering surfaces using microstructurally-sensitive crystal plasticity models to shed light on the mechanisms responsible for galling.

It is therefore clear that the deformation of contacting surfaces at elevated temperatures is important in assessing the propensity of a surface to galling. Although Barzdajn et al. [31] have introduced a new mechanistic modelling methodology for assessing the potential for galling in 316L steel at 20°C, there has so far been limited mechanistic assessment of thermal effects (frictional and internal plastic heat generation; elevated temperature) and role of adhesion in the galling of hard facing alloys. Another area which has not been thoroughly assessed is the role played by the strength of adhesion at this micron length scale, or in relation to temperature. The assessment of these effects is the primary objective of this work.

Sliding surfaces potentially generate transient heating associated with friction, the flash temperature phenomenon [32]. Frictional heating has been investigated at the macroscale, often dealing with sheet metal forming conditions [33,34], very different to the high-load, short-displacement conditions relevant to galling. The internal heating associated with plasticity is almost universally neglected when considering sliding surfaces but has been shown to be significant during highly localised deformation in other systems, for example, the formation of adiabatic shear bands [35]. It might be expected that if the deformation of surface asperities is sufficiently localised and the rate of deformation appropriately high, then the heat produced during plastic deformation could soften the material, leading to more extensive deformation. This together with frictional heating, could enhance deformation and be involved in the initiation of galling. Since the deformation at this length scale has been shown to be complex, further complicated by the length scale of asperities being comparable to the grain size in many fine grained hot isostatically pressed (HIPed) materials, this remains an open question. Hence, a thorough mechanistic approach is required to assess the role of heat generation.

The effect of elevated temperature in galling is also important, although macroscopic hardness has been shown to be a poor indicator in this respect [10]. Materials undergo a reduction in both stiffness and critical resolved shear stress, and this could become significant under highly localised deformation. This, along with any thermal softening, could be consequential to deformation at elevated temperatures. The framework developed by Barzdajn et al. [31] has shown some agreement with experimental galling results and is utilised in the present work to address these thermal effects. In what follows, the techniques employed to describe the strength of adhesion, frictional heating, internal heat generation and transfer, and the crystal plasticity of contacting surfaces are described, together with a summary of the contact galling methodology adopted. The assessment of frictional heating, internal heat generation through plasticity, strength of adhesion and elevated temperature in galling response are then presented, and the predicted galling behaviour is assessed against a range of experimental data available in the literature.

## 2 Thermo-mechanical material behaviour

A grain-level approach was developed using the crystal plasticity finite element (CPFE) method. Elevated temperature material behaviour was included in the material model with coupled heat generation and conduction. The resulting model was implemented within ABAQUS [36]. Simulations were performed using the facilities of Imperial College Research Computing Service [37].

## 2.1 Conduction with internal heat generation

An energy balance yields a relationship between thermal conduction and a distributed heat generation source. Consider a volume d$V$, with surface d$S$, net heat flux $\boldsymbol{J}$ leaving the volume, arbitrary volumetric heat generation rate $\dot{Q}_{\text{gen}}$, mass density $\rho$ and internal energy accumulation $u$. A heat balance over this volume leads to

$$\int_V \dot{Q}_{\text{gen}} \, dV = \int_S \boldsymbol{J} \cdot d\boldsymbol{S} + \int_V \rho \frac{\partial u}{\partial t} \, dV \tag{2.1.1}$$

Applying Fourier's law, the divergence theorem and writing the accumulation of internal energy in terms of temperature $T$, where materials properties in general can be specified as functions of temperature, gives

$$\dot{Q}_{\text{gen}} = -\lambda \nabla^2 T + \rho c_p \frac{\partial T}{\partial t} \tag{2.1.2}$$

where $c_p$ is the specific heat capacity and $\lambda$ the thermal conductivity. No conditions on the source term $\dot{Q}_{\text{gen}}$ have yet been prescribed and in this case is specified as the heat due to plastic deformation.

## 2.2 Kinematics and crystal plasticity

The crystal plasticity approach described in [38] is applied in this study, with inclusion of anisothermal effects. The deformation gradient tensor $\mathbf{F}$ is multiplicatively decomposed into elastic, plastic and thermal parts, $\mathbf{F}^e$, $\mathbf{F}^p$ and $\mathbf{F}^\theta$ respectively [35,39].

$$\mathbf{F} = \mathbf{F}^e \, \mathbf{F}^p \, \mathbf{F}^\theta \tag{2.2.1}$$

The thermal deformation gradient relates the undeformed configuration to the thermally expanded, unstressed configuration. This evolves with temperature, where $T_0$ is the temperature prior to any deformation and prescribed as an initial condition [39].

$$\dot{\mathbf{F}}^\theta \, \mathbf{F}^{\theta^{-1}} = \dot{T} \boldsymbol{\alpha} \quad \text{where} \quad \boldsymbol{F}^\theta = \boldsymbol{0} \quad \text{for} \quad T = T_0 \tag{2.2.2}$$

$$\boldsymbol{\alpha} = \alpha_{ij} \, \boldsymbol{a}_i \otimes \boldsymbol{a}_i \tag{2.2.3}$$

The thermal expansivity is given by tensor $\boldsymbol{\alpha}$, which has components $\alpha_{ij}$ with respect to basis vectors $\boldsymbol{a}_i$ aligned with the reference configuration of the crystal. For the face-centred cubic system, $\boldsymbol{a}_i$ are aligned with the <100> directions with $\boldsymbol{a}_1 \parallel$ [100] and so on.

Slip is assumed to be the sole deformation mechanism. $\mathbf{F}^p$ is a function of the crystallographic shear rate $\dot{\gamma}$ and its associated slip plane normal $\boldsymbol{n}$ and direction $\boldsymbol{s}$. The total slip contribution is found through summation over all systems (each denoted by $\kappa$) [38].

$$\dot{\mathbf{F}}^p = \mathbf{L}^p \, \mathbf{F}^p \tag{2.2.4}$$

$$\mathbf{L}^p = \sum_{n=1}^{\kappa} \dot{\gamma}^\kappa \left( \boldsymbol{s}^\kappa \otimes \boldsymbol{n}^\kappa \right) \tag{2.2.5}$$

A physically based slip rule [38] is used to quantify crystallographic shear, driven by dislocation glide with contributions from pinning and thermally activated escape

$$\dot{\gamma}^\kappa(\tau^\kappa) = \rho_\mathrm{m} \, f \, |\boldsymbol{b}^\kappa|^2 \exp\left(-\frac{\Delta F}{kT}\right) \sinh\left(\frac{(\tau^\kappa - \tau_c^\kappa)\,\Delta V}{kT}\right) \tag{2.2.6}$$

with $\rho_\mathrm{m}$ the density of mobile dislocations, $f$ the frequency of dislocation escape events, $\boldsymbol{b}^\kappa$ the burgers vector, $\Delta F$ the activation energy, $k$ the Boltzmann constant, $T$ the temperature, $\tau^\kappa$ the resolved shear stress on slip system $\kappa$, $\tau_c^\kappa$ the critical resolved shear stress, and $\Delta V$ the activation volume.

Hardening is accounted for by the local accumulation of slip. The statistically stored dislocation (SSD) density is taken to evolve incrementally with time such that

$$\rho_\mathrm{SSD}(t + \Delta t) = \rho_\mathrm{SSD}(t) + \gamma' \, \dot{p} \, \Delta t \tag{2.2.7}$$

with

$$\dot{p} = \sqrt{\frac{2}{3} \, \mathbf{D}^p : \mathbf{D}^p} \tag{2.2.8}$$

where $\gamma'$ is the hardening coefficient and $\mathbf{D}^p$ is the plastic deformation rate tensor. The slip strength is updated with evolving SSD density.

$$\tau_c^\kappa(T) = \tau_{c,\text{initial}}^\kappa(T) + G(T)\,|\boldsymbol{b}^\kappa|\,\sqrt{\rho_{\text{SSD}}} \tag{2.2.9}$$

The temperature sensitivities of the critical resolved shear stress and shear modulus $G$ are explicitly represented, as detailed in Section 2.3.

Schmid's rule is used to calculate the resolved shear stress $\tau^\kappa$ for a given slip system under a local stress state $\boldsymbol{\sigma}$ and to determine the active slip systems (those where the slip strength is exceeded).

$$\tau^\kappa = \boldsymbol{\sigma}\,\boldsymbol{n}^\kappa \cdot \boldsymbol{s}^\kappa \tag{2.2.10}$$

For anisothermal elevated temperature simulations, these equations are simultaneously solved alongside those for the heat balance to capture the temperature sensitivity of the material properties.

The properties within the slip rule (Table 1) were assumed to be constant with temperature and were those determined during the calibration study performed in [31] obtained from uniaxial tensile testing data [40].

Table 1: Slip rule parameters.

| Property | Unit | Value |
|---|---|---|
| Mobile dislocation density, $\rho_m$ | m$^{-2}$ | $9.69 \times 10^9$ |
| Burgers vectors magnitude, $|\boldsymbol{b}|$ | m | $2.54 \times 10^{-10}$ |
| Attempt frequency, $f$ | s$^{-1}$ | $1.0 \times 10^{11}$ |
| Activation energy, $\Delta F$ | J | $2.6 \times 10^{-20}$ |
| Boltzmann constant, $k$ | J K$^{-1}$ | $1.38 \times 10^{-23}$ |
| Activation volume, $\Delta V$ | m$^3$ | $25.2\,|\boldsymbol{b}|^3$ |
| Hardening coefficient, $\gamma'$ | m$^{-2}$ | $7.0 \times 10^{13}$ |

## 2.3  Temperature dependent material properties

Due to the scarcity of data for the mechanical behaviour of iron-based hard facings, 316L stainless steel is considered due to its well characterised properties over a wide range of temperature; 316L is not too dissimilar to some iron-based hard facings such as Nitronic 60 [41], and experimental galling data are available. The temperature sensitivities of property data were taken from the literature, ensuring compatibility with the 20°C values used in the earlier work of Barzdajn et al. [31], with resulting relationships given in Table 2. Anisotropic elastic properties were obtained from [42].

Table 2: Material properties as functions of absolute temperature $T$.

| Property | Unit | Equation | Source |
|---|---|---|---|
| Thermal conductivity | W m$^{-1}$ K$^{-1}$ | $\lambda = 0.0125\,T + 11.3$ | [43] |
| Specific heat capacity | J kg$^{-1}$ K$^{-1}$ | $C_p = 0.0879\,T + 467.3$ | [43] |
| Thermal expansivity | K$^{-1}$ | $\alpha = (0.0023\,T + 12.34) \times 10^{-6}$ | [43] |
| Density | kg m$^{-3}$ | $\rho = -0.428\,T + 7974$ | [43] |
| Young's modulus | GPa | $E = -0.0737\,T + 119.1$ | [43] |
| Shear modulus | GPa | $G = -0.0385\,T + 136.3$ | [44] |
| Critical resolved shear stress | MPa | $\tau_c = 0.00019\,T^2 - 0.2774\,T + 144.1$ | [45] |

## 2.4 Internal heat generation associated with plastic working

During deformation, the work done comprises recoverable elastic work and dissipated plastic work (neglecting dislocation interaction energy). The total plastic work $W_p$ manifests itself as strain energy stored within the stress fields associated with dislocations or as heat [46–48]. The rate of plastic work, $\dot{W}_p$, is expressed as the product of the effective plastic strain rate $\dot{p}$ and the von Mises stress $\sigma_{\text{VM}}$.

$$\sigma_{\text{VM}} = \sqrt{\frac{3}{2}\boldsymbol{\sigma}':\boldsymbol{\sigma}'} \tag{2.4.1}$$

$$\dot{W}_{\text{p}} = \dot{p}\,\sigma_{\text{VM}} \tag{2.4.2}$$

where $\boldsymbol{\sigma}'$ is the deviatoric stress tensor. The rate of plastic heat generation, $\dot{Q}_{\text{p}}$, is related to the rate of plastic work by the factor $\beta$, taken as 0.95. This is consistent with most metals [46,47].

$$\dot{Q}_{\text{p}} = \beta\,\dot{W}_{\text{P}} = \beta\,\dot{p}\,\sigma_{\text{VM}} \tag{2.4.3}$$

## 2.5 Friction heat generation

The frictional force, $\boldsymbol{F}_{\text{fric}}$, developed between two surfaces is given by

$$\boldsymbol{F}_{\text{fric}} = -\mu\,|N|\,\frac{\boldsymbol{v}}{|\boldsymbol{v}|} \tag{2.5.1}$$

where $\mu$ is the dynamic friction coefficient, $N$ the normal force and $v$ the relative sliding velocity. The work rate per unit surface area can be expressed in terms of the surface shear stress $\tau_{surf}$ and the rate of slipping between the surfaces $\dot{\gamma}_{fric}$ (dimensions L T$^{-1}$).

$$\dot{W}_{fric} = \tau_{surf}\, \dot{\gamma}_{fric} \qquad (2.5.2)$$

All frictional work is assumed to manifest itself as a surface heat source in order to give a worse case estimate of the effects of frictional heating.

## 2.6 Adhesion modelling with friction

The surfaces modelled here approximate perfectly clean, contaminant free, bare metal surfaces and would be expected to strongly adhere to one another. This may be captured in the model through a non-linear frictional approach. The friction coefficient $\mu$ represents the contribution of asperity interactions below the length scale considered in this model. The frictional model as discussed in Section 2.5 allows for arbitrarily high frictional shear stresses at the interfaces between the contacting surfaces. In reality, adhered junctions can fail, either through plastic deformation or fracture.

Equation 2.5.1 can be expressed in terms of shear stress $\tau$ and contact pressure $p$.

$$\tau = -\mu\, p\, \frac{\boldsymbol{v}}{|\boldsymbol{v}|} \qquad (2.6.1)$$

A sensible shear stress limit $\tau_{max}$ would be the critical resolved shear stress. Once the shear stress reaches this value then the material would yield, and plastic slip would occur. This behaviour is shown in Figure 1.

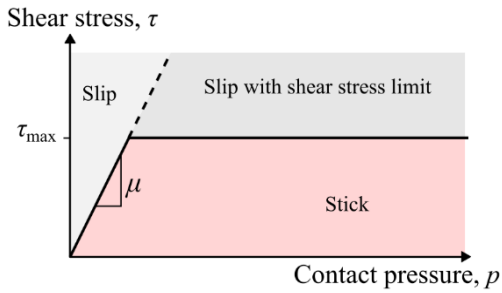

Figure 1: Frictional model used to describe adhesion, with the friction coefficient $\mu$ and the shear stress limit $\tau_{max}$. The dashed line represents Coulomb friction without a shear stress limit.

This model assumes that the surfaces adhere with a well-defined, uniform shear strength. In true materials, this will depend on several factors: the crystal structure, the crystallographic plane at the surface, the misorientation between the surfaces, surface contamination, and the temperature [24,29,49]. As such, the critical resolved shear stress is likely to be an overestimate of the strength of these junctions and would give a worst-case scenario (in terms of severity of plastic deformation) for the stress state at the adhered junctions [50].

## 3 The representative galling model

This study builds on the methodology developed by Barzdajn et al. [31], but including the effects of transient heat conduction, adhesion at contacting asperities, plastic and frictional heating and temperature-coupled material properties. The model is based upon the ASTM G196 testing procedure for galling resistance [16].

### 3.1 Model geometry and microstructure

To ensure that the appropriate length scale for galling was modelled, a small region of interest of the order of 10 μm was selected, as shown schematically in Figure 2. The curvature and rotational nature of the system is reasonably neglected at this length scale with linear displacement used instead for the sub-region considered.

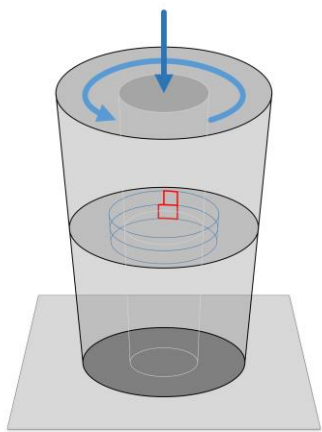

Figure 2: Schematic diagram of the region of interest selected from ASTM G196 geometry. The cylindrical slice is shown in blue whilst the modelled section is shown in red.

The model consisted of two parts, an upper punch section (Part 1) and a lower stationary section (Part 2), shown in Figure 3. A synthetic microstructure of 2.55 µm hexagonally shaped grains was used to represent the fine-grained microstructures seen in HIPed materials [51]. Both sections were polycrystalline (Figure 3(b)), single phase austenite with a random crystallographic texture (Figure 3(c)), generated with uniform random sampling. Surface profiles were generated from profilometry measurements of RR2450 [52], resulting in a surface roughness for these surfaces of $R_a = 0.1$ µm. Detail of the surfaces is shown in Figure 4.

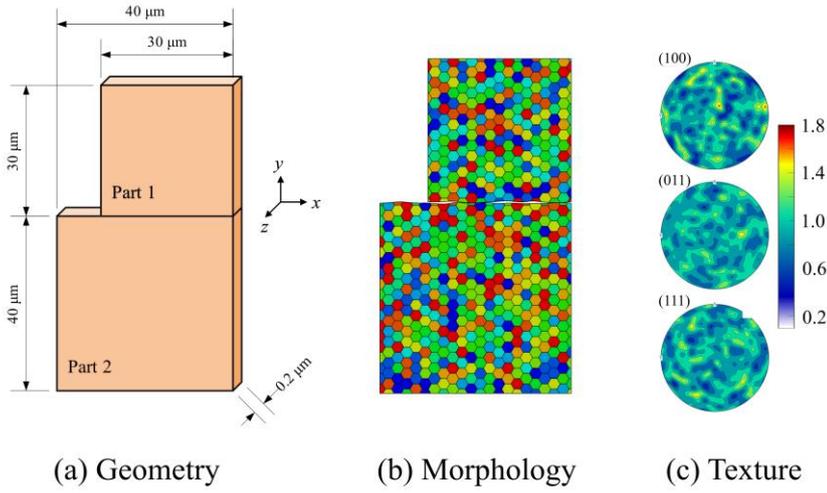

(a) Geometry     (b) Morphology     (c) Texture

Figure 3: Model geometry and morphology. The colours in (b) are to differentiate different grains but do not explicitly represent the crystallographic orientations. (c) shows plots of the orientation distribution function, visualised using the MTEX package [53].

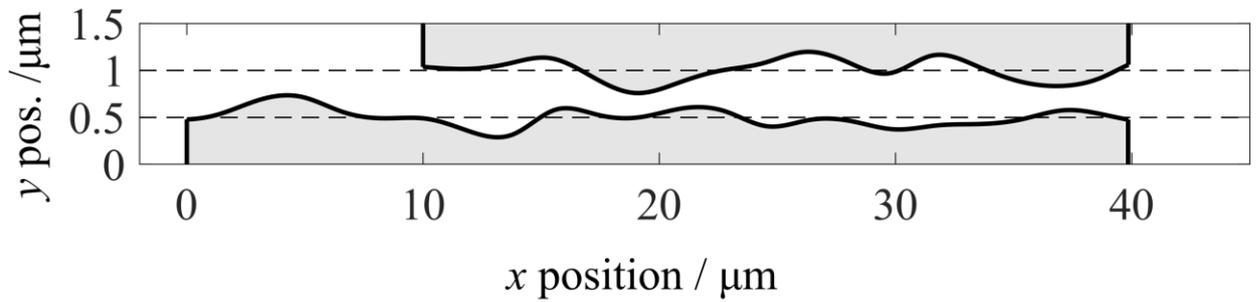

Figure 4: Surface detail showing the profile of the asperities. The dashed lines indicate the mean line of the surfaces. Note the difference in scale between the two axes.

A single layer of 20-noded hexahedral elements (C3D20RT) was applied, with areas likely to undergo plastic deformation assigned a refined mesh (0.2 μm refined element size, 0.4 μm otherwise). The 2D microstructure was extruded through a depth of 0.2 μm, resulting in prismatic grains. A plane strain condition was imposed to reflect the thin layer represented in Figure 3.

### 3.2 Galling process model

A representative sliding time was required to capture the transient nature of heat conduction, with a sliding velocity of 1 mm s$^{-1}$ was used throughout. This sliding speed is representative of that found in both in-service valves and ASTM G196 [16]. Simulations were performed in three steps, as illustrated in Figure 5.

1. Contact: Part 1 is lowered under displacement control until it contacts Part 2
2. Loading: Part 1 is loaded normally and monotonically to the desired maximum load
3. Sliding: Part 1 is displaced 10 µm horizontally. The vertical loading remains applied.

For step (3), a sliding time of 0.01 s (sliding distance 10 µm) gave the representative 1 mm s$^{-1}$ velocity.

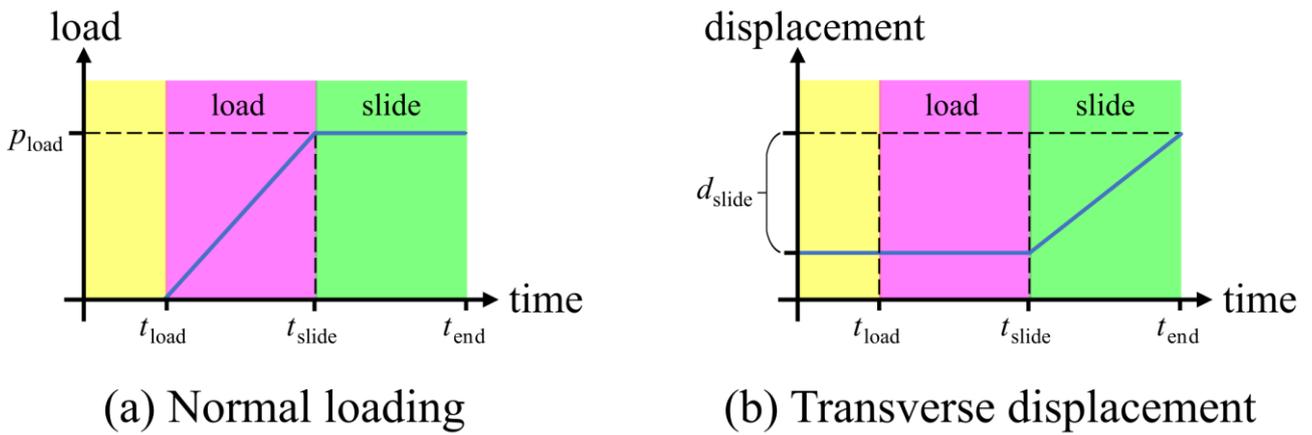

Figure 5: The evolution of loading and displacement conditions with time. The yellow section denotes contact (Step 1), the pink loading (Step 2) and the green sliding (Step 3). The timescales shown are not to scale.

The mechanical boundary conditions imposed are shown in Figure 6(a). All displacements on the bottom surface of Part 2 and all out of plane ($z$ – direction) displacements were constrained. During the loading step, a normal load was applied with a monotonically increasing magnitude up to $p_{load}$. Sliding was controlled by displacing the upper surface of Part 1 a distance $d_{slide}$ in the negative $x$ - direction whilst maintaining the normal load. All external surfaces of the model were taken to be adiabatic since any heat generated local to the contacting surfaces is unlikely to escape from the body and therefore the heat flux at the edge of the region would be negligible.

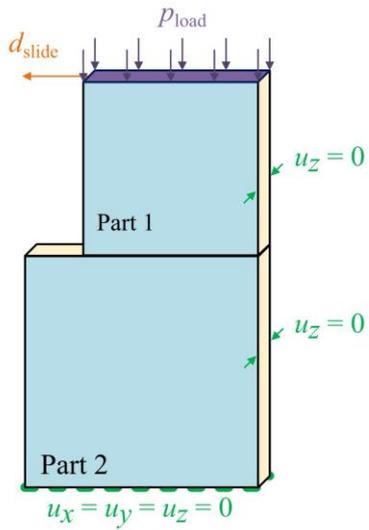 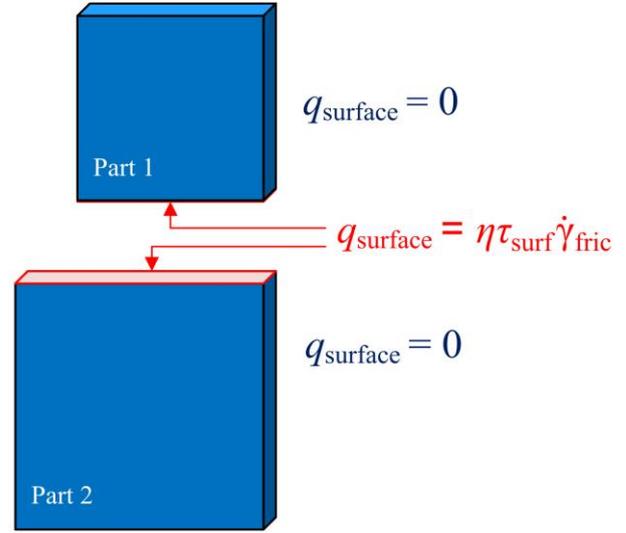

(a) Mechanical boundary conditions  (b) Thermal boundary conditions

Figure 6: Schematic representations of the model boundary conditions. In (a), the blue front and back surfaces are constrained in the z direction whilst the displacements in all directions are constrained on the base of Part 2. In (b), all blue surfaces are adiabatic. Heat fluxes associated with friction are applied on the red surfaces, but no heat transfer is permitted between the two bodies.

The frictional heat generation was accounted for as a moving surface source. The heat flux was calculated using Equation (3.2.1). The factor $\eta$ controls the distribution of heat between the two surfaces, taken as 0.5 throughout.

$$q|_{\text{surf}} = \begin{cases} \eta \tau_{\text{surf}} \dot{\gamma}_{\text{fric}}, & \tau_{\text{surf}} > 0 \\ 0, & \tau_{\text{surf}} = 0 \end{cases} \quad (3.2.1)$$

Heat transfer between the surfaces was assumed to be negligible when compared with the heat flux due to friction. Thermal boundary conditions are shown schematically in Figure 6(b). For models not examining the strength of adhesion, a friction coefficient of 0.1 was used to represent friction due to surface interactions at length scales below that explicitly represented in the geometric model.

## 3.3 Galling quantification

The methodology of Barzdajn et al. [31] is used to relate the CPFE model to macroscopic galling frequencies and is discussed here briefly for completeness. The extent of plastic deformation caused by sliding was quantified by the plastic reach $p_R(l)$ defined as

$$p_R(l) \propto \iiint_V p_{\text{eff}}(x,y,z,l)\, d(x,y,z)\, dx\, dy\, dz \qquad (3.3.1)$$

where $p_{\text{eff}}(x,y,z,l)$ is the effective plastic strain at position $(x,y,z)$ and $d(x,y,z)$ is the depth of position $(x,y,z)$ below the material surface. This product was integrated over the plastic zone $V$.

The characteristic galling load $l_0$, the load at which galling occurs, was defined as the microscopic yield point of the first asperity to plastically deform, such that the local effect plastic strain exceeds 0.2%.

$$l_0 = l(\max(p_{\text{eff}}(x,y,z,l)) \geq 0.002) \qquad (3.3.2)$$

The plastic reach was found to be well approximated by a power-law relationship for loads exceeding $l_0$, where $\xi$ is a constant.

$$p_R(l) \propto \left(\frac{l}{l_0}\right)^\beta \Rightarrow \ln p_R(l) = \beta \ln l - \ln \xi - \beta \ln l_0. \qquad (3.3.3)$$

By relating the cumulative hazard rate $H$ to the computed plastic reach, the power-law parameters can be evaluated from multiple simulations at various loads, allowing the shape factor for the distribution to be quantified.

$$H(l) = \xi p_R(l) = \left(\frac{l}{l_0}\right)^\beta. \qquad (3.3.4)$$

For the Weibull distribution, the galling frequency $F$ (cumulative probability of galling for loads less than or equal to a given load $l$), is related to the cumulative hazard rate by

$$F(l) = 1 - \exp(-H(l)). \qquad (3.3.5)$$

Finding the characteristic galling load $l_0$ and the shape factor $\beta$ gives the galling frequency as a function of applied load.

$$F(l) = 1 - \exp\left(-\left(\frac{l}{l_0}\right)^\beta\right) \qquad (3.3.6)$$

# 4  Investigation of internal heat generation

The two sources of internal heat generation (plasticity and friction) were first examined separately and then their combined effect was assessed. The rate of heat conduction was much higher than the rate of heat generation and no regions of localised, persistent heating were observed, even in the presence of both generation mechanisms. This was verified by performing simulations with a time step of $10^{-5}$ s (c.f. 0.01s sliding time) to capture any rapid heating. Heat sources are seen in Figure 7, with heating localised to asperities in contact, undergoing deformation.

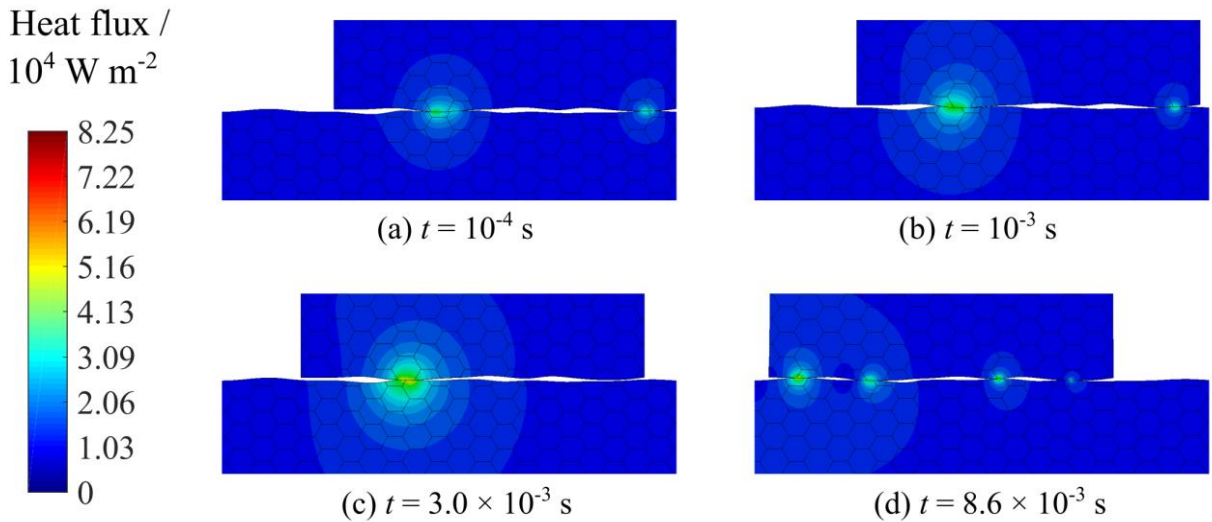

Figure 7: Contour plots of heat flux (W m$^{-2}$) during the sliding simulation step. Both frictional and plastic heating were active, and the applied load was 50 MPa. The total sliding time was 0.01 s.

The rapid heat conduction is expected since 316L stainless steel is an excellent conductor of heat. It was perhaps unexpected that the rate of deformation (and therefore heat generation) was much lower than that of conduction, preventing any temperature localisation or local softening.

Near uniform temperatures were seen throughout deformation and temperature contours are shown in Figure 8. The temperature differential seen between the two parts was due to the heat transfer across the contact surface being neglected. This is of little consequence in light of the negligible local temperature rises. The assumption that the slip rule parameters were temperature independent remains valid in light of the negligible increases in temperature.

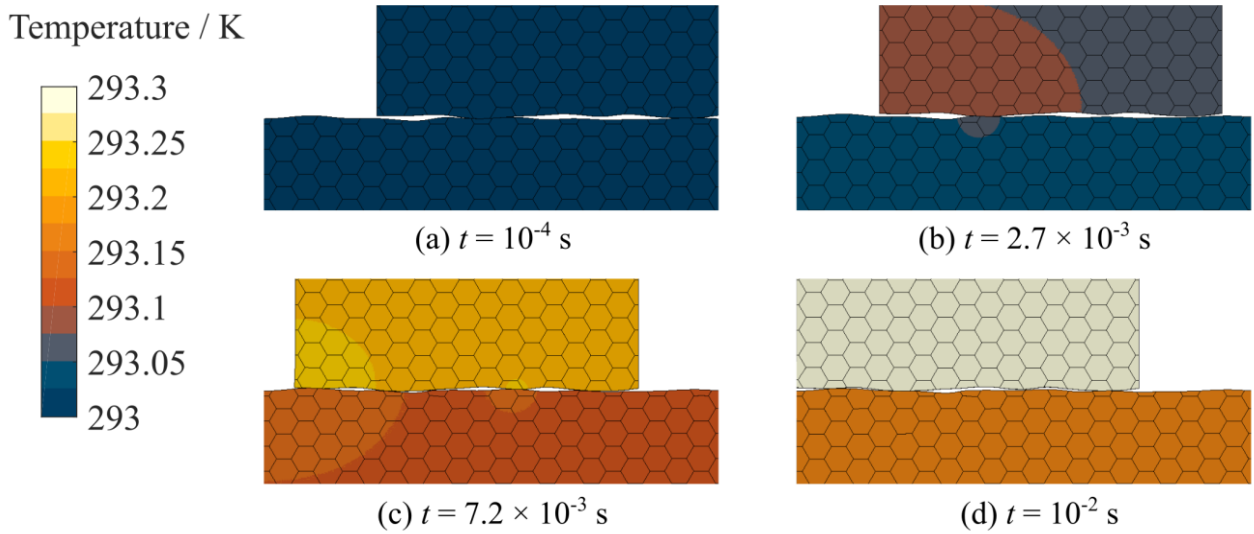

Figure 8: Temperature (K) contour plots during sliding (total time 0.01 s, 50 MPa applied load, friction coefficient 0.1 without shear stress limit). At the end of sliding (d), the upper and lower parts have peak temperatures of 293.3 K and 293.2 K respectively.

The modest temperature rises are accounted for by the rate of conduction far exceeding the rate of sliding. Studies considering macroscopic frictional heat generation show much larger temperature rises. In a metal forming context, Gåård et al. [24] studied sliding velocities in excess of 100 mm s$^{-1}$, over 100 times that considered here, achieving temperature rises of the order of 100 K.

Both the characteristic galling load and shape factor were insensitive to heat generation. Since $l_0$ is determined under normal loading and at the yield point of the material, heating cannot occur before this point. The magnitude of heat generation was insufficient to influence the plastic deformation and therefore $\beta$. Therefore, galling frequency and localised asperity deformation was found to be insensitive to internal heat generation.

## 5  Investigation of adhesion model

To assess the sensitivity of the galling response to adhesion, combinations of friction coefficient $\mu = 0.1$, 0.5, 1.0 and $\tau_{max} = 80$, 160 and no shear stress limit were applied under normal loads of 10, 30 and 50 MPa. Simulations under the more severe loading conditions ($\mu = 0.5$, 1.0 with no shear stress limit) proved difficult to converge. This lack of convergence was likely due to the large levels of plasticity due to the non-physical strengths of the adhered junctions, with these junctions displaying non-physical levels of shear stress.

Minimal changes in the deformation behaviour between the differing adhesion models were observed, with the geometry of the surfaces appearing to control the plastic deformation of the asperities. Changes in surface profile after sliding and true contact area during sliding were almost entirely insensitive to changes in adhesion strength. This was due to the interfaces being unable to support sufficient shear force to cause enhanced deformation.

For all friction coefficients, the shear stress limit controlled the horizontal force required to satisfy the displacement constraint. Figure 9(a) shows the reduction in the horizontal force with the reduction in the shear stress limit, under a 50 MPa normal load. The upper line represents the friction model without any shear stress limit, equivalent to strongly adhered surfaces. The data for $\tau_{max}$ = 160 MPa show a slight reduction in horizontal force, indicating that the surface shear stresses are typically below $\tau_{max}$ for much of the sliding. Data for $\tau_{max}$ = 80 MPa show considerable reductions in force, indicating extensive sliding of the surface. Figure 9(b) shows an increase in horizontal force for the $\tau_{max}$ = 160 MPa data, suggesting that the higher shear stress limit is not reached for all contacting areas for $\mu$ = 0.1.

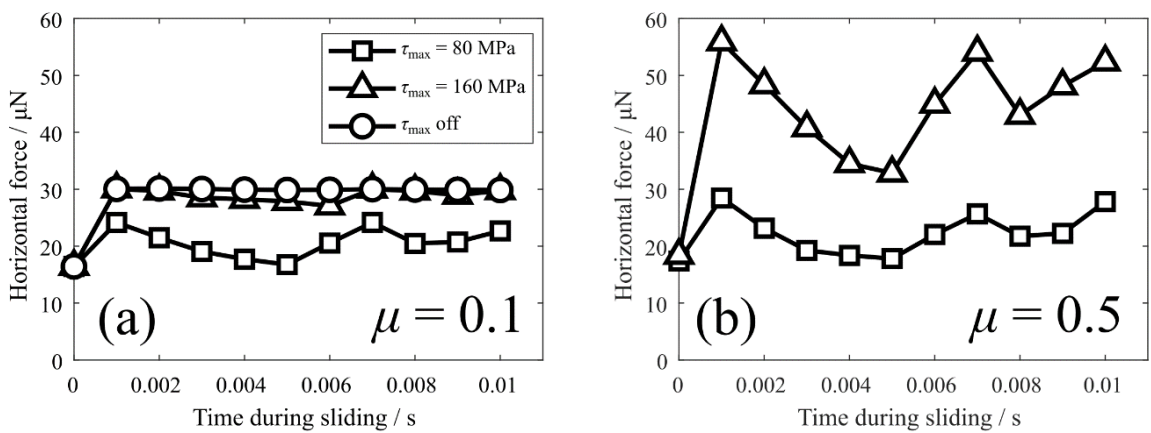

Figure 9: Horizontal force applied to upper model part with friction coefficients $\mu$ = 0.1 (a) and $\mu$ = 0.5 (b). Data for $\mu$ = 1.0 were identical to that for $\mu$ = 0.5. Note the absence of data in (b) for $\tau_{max} \to \infty$ due to convergence issues preventing a solution.

Frictional behaviour was controlled by the shear stress limit, with little dependence on the friction coefficient found. For all but the lowest friction coefficient, $\tau_{max}$ can be easily achieved, due to the small true contact area causing large stress concentration and localisation. Negligible differences in the deformation behaviour were found with no substantial dependence on adhesion (frictional model) observed. For all models, substantial plastic deformation was seen but this was due to the geometry of the asperities. The local stress states surrounding asperities were largely compressive with only a small shear component. Normal stresses exceeding several 100 MPa were typical, often around the 800 MPa mark for this particular surface pairing, far outweighing the adhesive shear component.

Mechanistically, this shows that the absolute strength of adhesion is inconsequential to the deformation of asperities at this length scale. Whilst this model does not fully capture the fracture of adhered surfaces, it does show that, under sliding, the asperity interfaces would slip (either through crystallographic slip leading to plastic deformation or decohesion).

The characteristic galling load was insensitive to adhesion strength, as would be expected since $l_0$ is determined under normal loading. The Weibull shape factor $\beta$ was also insensitive to friction model, since changes in deformation for differing adhesion strengths were mild. As such, galling frequency was found to be insensitive to adhesion strength.

The surfaces considered in the present study approximate contaminant free, bare metal surfaces, analogous to surface after surface layers have been scoured away by previous motion. This suggests that the strength of adhesion is irrelevant after the onset of metal-on-metal contact and that surface geometry controls the deformation. This further suggests that a change in the deformation mechanism of the surfaces controls galling, whether this is phase transformation [8,11,14,19,20,26] or removal of protective surface oxide layers [8,14,19,26,28]. Deformation at this length scale therefore appears to be controlled by the geometry of the surfaces rather than the adhesion between them, in bare-metal single phase austenitic materials.

# 6 Investigation of elevated temperature galling

## 6.1 Initial investigations with low roughness surfaces

The effects of internal heat generation are assumed negligible for all subsequent analyses, which are hence performed isothermally. Elevated temperature analyses were performed isothermally at 25°C, 100°C, 200°C and 300°C. A single surface profile and microstructure was used throughout and subjected to normal loads of 10, 20, 30 and 40 MPa. A simple frictional model with no shear stress limit and $\mu = 0.1$.

For increasing temperatures, the peak stress developed within the material during normally-loaded sliding contact decreased, albeit several times higher than the apparent applied stress for all temperatures. Contours of Mises stress (Figure 10) show both a reduction in peak stress and a slight change in stress profile. The softening of the material at elevated temperature increased the extent of plastic deformation, increasing the true contact area and distributing the load.

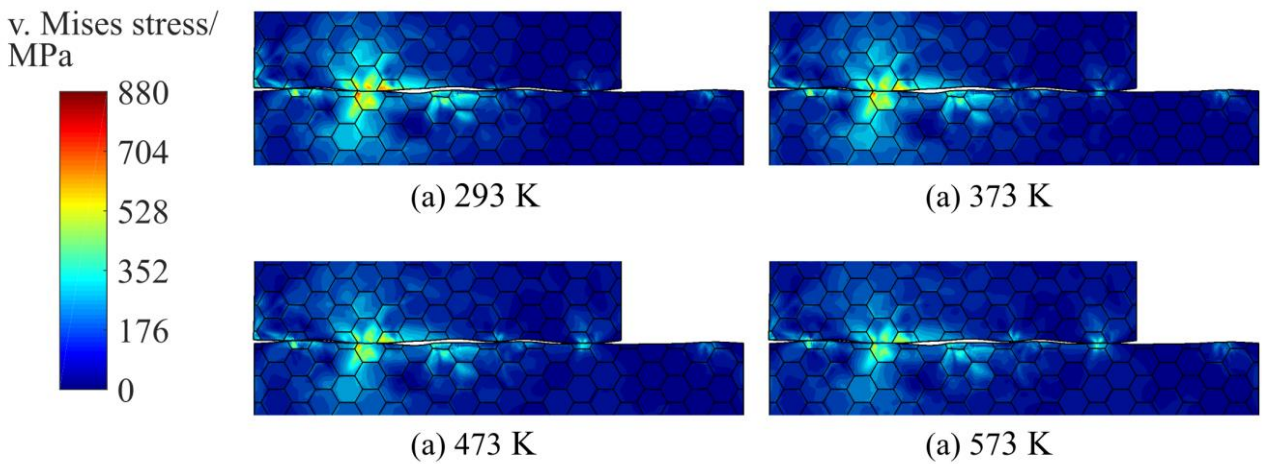

Figure 10: Contours of Mises stress at the end of sliding for an applied load of 40 MPa. As the temperature increases, the peak stress is reduced (note the absences of red areas) whilst new areas of stress appear due to an increase in asperity contact. The peak stresses are of the order of several hundred MPa for an applied load of 40 MPa.

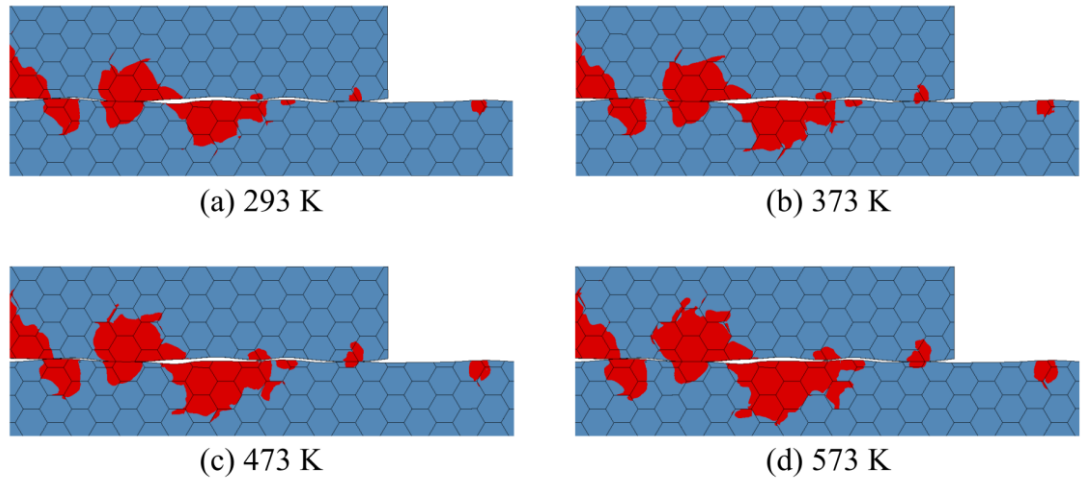

Figure 11: Plastic zones at the end of sliding with red and blue showing the yielded and unyielded areas respectively. The material is considered to have yielded when the effective plastic strain $p_{\text{eff}}$ has exceeded 0.2%. All four plots use a 40 MPa applied load.

The extent of plasticity increased with increasing applied load and temperature. Clear increases in the plastic zones (areas where $p_{\text{eff}}$ exceeded 0.2%) were seen with increasing temperature, demonstrated in Figure 11 and Figure 12(b). The effective plastic strain (averaged over the plastic zone) showed a strong dependency on temperature, with Figure 12(a) showing decreases in plastic strains with increasing temperature for all loads. An increase in $p_{\text{eff}}$ was seen between 10 and 20 MPa, with the effect saturating for further increases in load. The penetration depth of the plastic zone also increased with increasing temperature (Figure 12(c)). The plastic reach (Figure 12(d)) showed a substantial increase with temperature, approximately doubling in magnitude for each load for a temperature increase of 300 K.

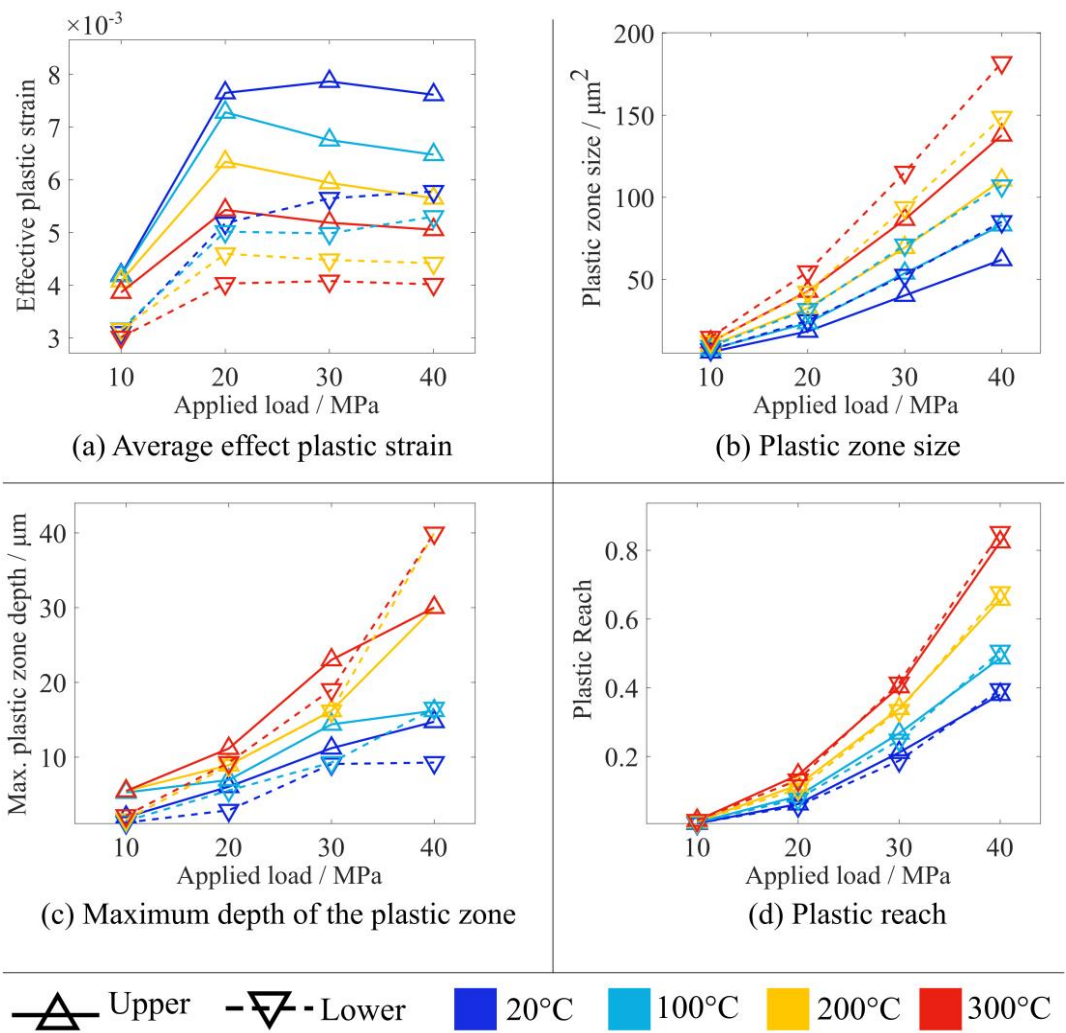

Figure 12: Various metrics of plastic deformation for a range of applied loads and temperatures. All values are those at the end of sliding.

The reduction in the average effective plastic strain can be rationalised by considering the deformation of asperities. At higher temperatures, the increased extent of plastic deformation increased the number of asperity contacts and hence the contact area. The increase in contact area reduces local stresses, resulting in a reaction in plastic strain.

The plastic reach (a function of effective plastic strain and its distance from the surface) showed both load and temperature sensitivity. To accommodate deformation, yielded material can further plastically deform, hardening in the process, or the plastic zone can expand. At higher temperatures, a lower stress was required to initial plastic deformation, and therefore a larger body of material yields for a given stress, expanding the plastic zone. This effect dominated the plastic reach, leading to an increase with increasing temperature.

Both characteristic galling load $l_0$ and Weibull shape $\beta$ parameter showed temperature sensitivity. The location of the microscopic yield point did not change with changing temperature. From room temperature to 300°C, $l_0$ decreased from 7.3 to 4.6 MPa as shown in Figure 13(a). The magnitude of these changes in $\beta$ was much smaller than those seen in $l_0$, with a reduction in the mean value of $\beta$ for upper and lower parts from 3.13 to 2.97 over the same temperature range (Figure 13(b)).

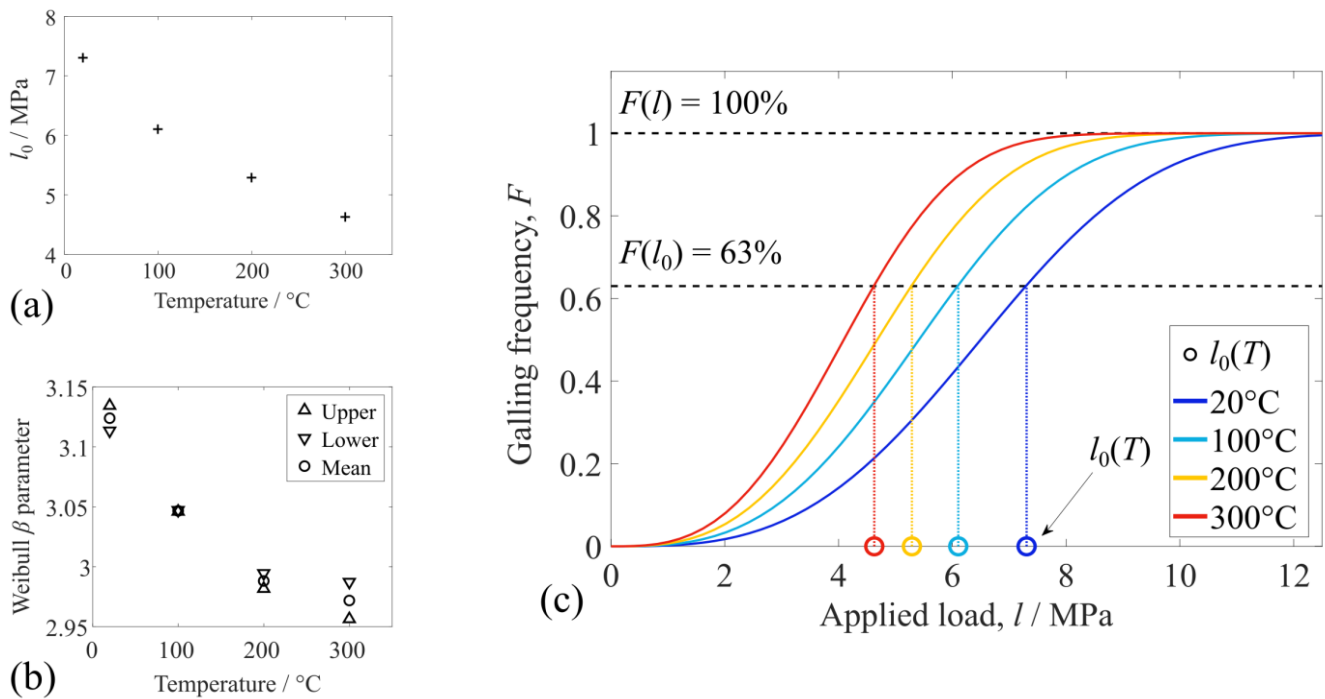

Figure 13: Galling frequency curves. (a) and (b) show the variation in $l_0$ and $\beta$ with temperature respectively. (c) shows the resulting galling frequency curves with temperature.

To understand the galling frequency response as a function of temperature, galling frequency curves were plotted (Figure 13(c)) at the investigated temperatures (Equation (3.3.6)). The reduction in characteristic galling load resulted in the galling frequency curve shifting to lower applied loads. The maximum gradient increased with increasing temperature, implying an increase in load sensitivity at higher temperatures (i.e. a fixed change in applied load has a greater influence on galling frequency at higher temperatures). The applied increase in temperature appears only to affect the characteristic galling load but not the Weibull shape factor. The changes found in the value of *β* are not large enough to impact the resulting galling frequency curve. From 20°C to 300°C, a reduction of 37% in $l_0$ was found (Figure 13(a)), with a reduction in slip strength of 39% over this range. A linear relationship between slip strength and $l_0$ was found, suggesting that the slip strength controls the temperature sensitivity of $l_0$.

The galling frequency behaviour of 316L at elevated temperature therefore appears to be controlled almost entirely by changes in characteristic galling load with the Weibull shape factor being relatively insensitive to temperature. Whilst reductions in characteristic galling load were observed, these were all relatively smooth, without the abrupt transition in galling resistance as reported by several investigators [8,11]. 316L is well known for its poor galling resistance at loads between 4 – 8 MPa [3], reducing by several MPa with increasing temperature [13,15,16,18,54,55]. These loads are much less than the typical galling loads for hard facings; for example Cockeram et al. [55] reported a threshold galling stress of 650 MPa for NOREM 02 and Stellite 6 exceeding 1200 MPa. Qualitatively, the results presented here are in agreement with the high temperature reduction in galling load.

## *6.2 Assessment of model performance against literature results*

The present study has thus far considered a single surface profile with a low surface roughness. To provide a more representative discussion and comparison with the galling results in the literature, surface profile properties (roughness) have been examined simultaneously with temperature.

### *6.2.1 Surface profile properties*

The more recent standard test, G196, has not found widespread adoption due to the large quantity of material and testing required and the associated expense [56]. As such, G196 results are not widely available in the literature, with G98 often favoured in spite of its qualitative nature. The data collection method presented in G196 does allow for comparisons to be made with the representative models investigated here [16].

---

[3] Results for both threshold galling stress as in ASTM G98 (ASTM International, 2009) and galling$_{50}$ in ASTM G196 (ASTM International, 2016). Although the two results are not directly comparable, they do give some insight into the loads required to initiate galling.

Harsha et al. [17,18] have recently implemented the G196 testing procedure with the addition of heating coils to allow galling frequency data to be recorded at elevated temperature, performing tests for various stainless steels at room temperature and 300°C (albeit deviating from the ASTM standard by rotating the sample twice rather than once). Significant reductions in galling resistance were found, concordant with the generally found galling behaviour at elevated temperature. An exact value for room temperature was not given in [18] so was taken as 25°C for our study.

Exact surface profile data are often not recorded. A single surface specification is given for the G196 as used in [18], namely that the arithmetic roughness is to be between 0.25 and 0.35 μm. Arithmetic surface roughness, $R_a$, is defined as

$$R_a = \frac{1}{N}\sum_{i=1}^{N}|y_i - \bar{y}| \quad \text{where} \quad \bar{y} = \frac{1}{N}\sum_{i=1}^{N} y_i \qquad (6.2.1)$$

for a surface with $N$ sampling points where $y$ is the height of the surface and $\bar{y}$ the mean line.

The model introduced in Section 3.1 has an arithmetic surface roughness of approximately 0.1 μm, lower than the values for the material used by Harsha et al. [18], with $R_a$ in the range 0.25 – 0.35 μm). Whilst individual surface roughness parameters do not provide the details needed to represent the full surface profile information which has been shown to be of much more relevance to galling resistance [57], the use of arithmetic roughness does enable models to be generated for a qualitative comparison with results in the literature.

Models with roughness values of 0.25 μm and 0.35 μm were generated to provide bounds on galling frequency. The profile previously described was scaled to give the desired surface roughness whilst retaining the form of the surface. This process is shown schematically in Figure 14 with Equation (6.2.2) showing the relationship between scaled and unscaled y-coordinates.

$$y_i^{\text{new}}(x) = \bar{y} + \left(y_i^{\text{old}}(x) - \bar{y}\right) \frac{R_a^{\text{new}}}{R_a^{\text{old}}} \tag{6.2.2}$$

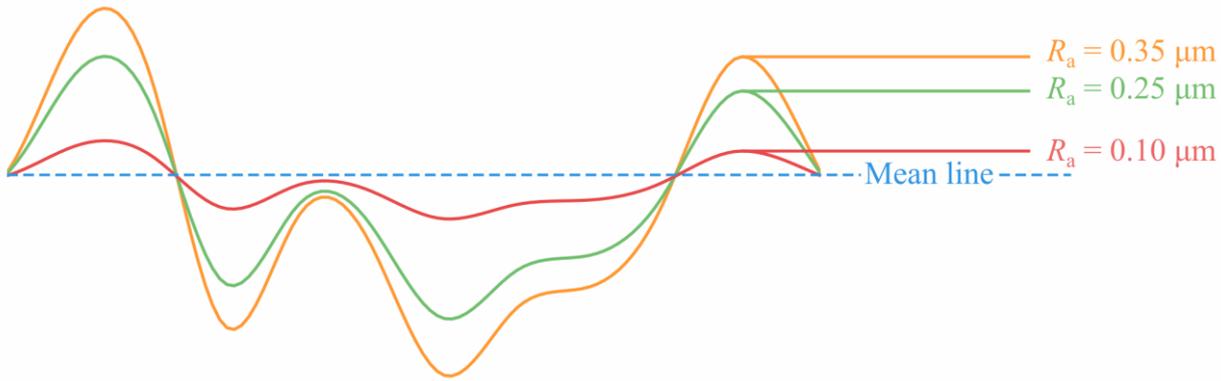

Figure 14: Representation of the surface scaling process. All three surfaces share a common mean line and profile but are scaled to give differing surface roughness values. These surface profiles are not to scale with the vertical height magnified for clarity.

Different arrangements of grains were required to accommodate different surface profiles and, accordingly, the orientations of individual grains varied between each realisation of the general model. The overall base texture and sampling technique remained unchanged but a unique texture and set of crystallographic orientations were used for each surface. Previous work has shown that galling resistance is relatively insensitive to crystallographic texture [57].

For a given nominal surface roughness, a wide range of possible surface profiles may be envisaged which satisfy the surface roughness constraint with very different surface geometries. This highlights the weaknesses inherent in the use of $R_a$ in quantifying surface profile. The wavelength of the profiles was quantified with the dimensionless ratio $\lambda/\lambda_0$, where $\lambda$ is the average separation between surface minima and $\lambda_0$ the grain diameter (2.55 μm). To vary the surface profiles, the wavelength was scaled by a stretch ratio of 1.5, as shown in Figure 15.

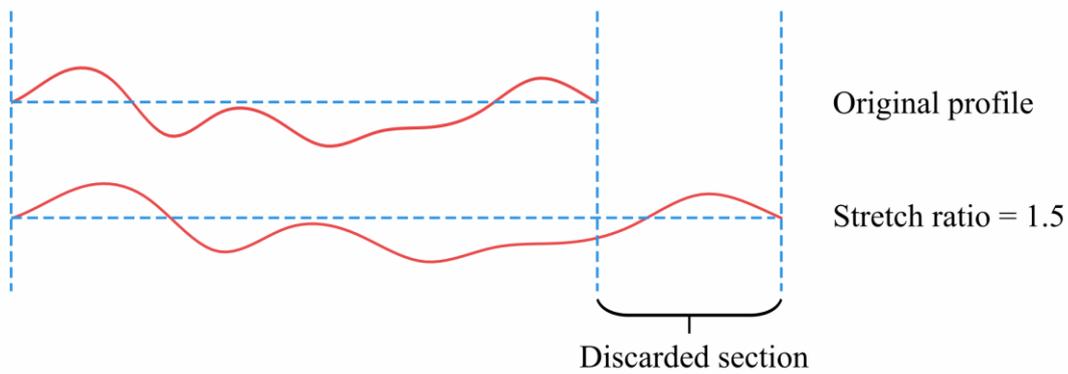

Figure 15: Schematic diagram detailing the horizontal scaling process, with original profile and the scaled profile after a stretch of 1.5. The discard section is removed in order for the profile to conform to the geometry of the model (30 and 40 μm side lengths for the upper and low sections of the model respectively).

Six different surface profiles were examined, combinations of three surface roughness values ($R_a$ = 0.1, 0.25, 0.35 μm) and two surface wavelengths ($\lambda/\lambda_0$ = 2.7, 4.1), at room temperature (25°C) and 300°C. This represents just a small selection of the possible surface profiles which would conform to these asperity spacing and surface roughness criteria. A more extensive investigation into the roles of surface profile and asperities spacing is out of the scope of this work. All surfaces share a common base surface profile on which scaling and stretching operations were applied.

### 6.2.2 Deformation behaviour

The particular surface geometry of the material surfaces plays a controlling role in the deformation of the surfaces. Scaling both the amplitude and wavelength of the profile caused significant changes to the surface geometry. This led to varying asperity interactions, changing the number of asperities in contact at any one time. Trends with temperature were broadly the same as those found during the low roughness investigations and, as such, this discussion will focus on surface profile sensitivity.

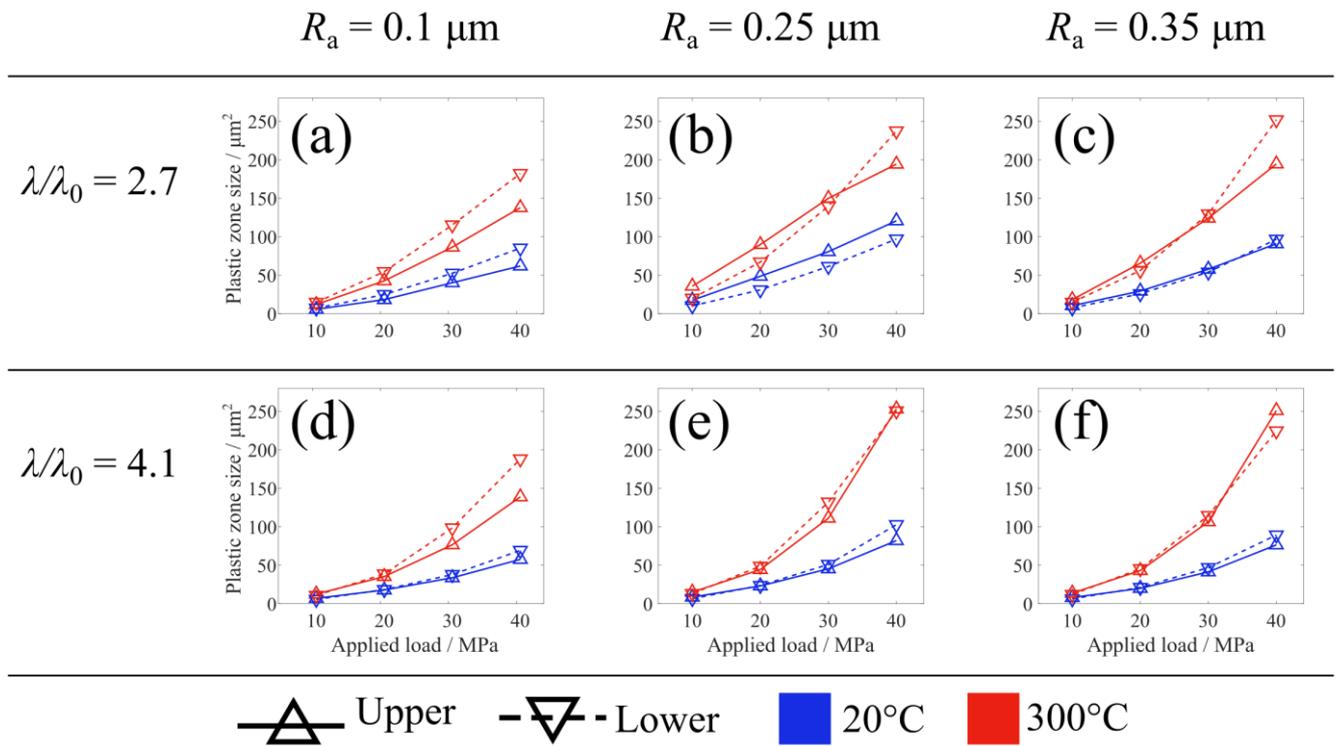

Figure 16: Plastic zone size for the six surfaces for the two temperatures considered. The plastic zone size in the lower region was found to be larger for all surfaces.

Surface roughness had a much stronger influence on plastic zone size than horizontal scaling, detailed in Figure 16. Asperities tended to graze over on another for low roughness ($R_a = 0.1$ μm). At the other extreme, high surface roughness prevented much of the surfaces from contacting, leading to deformation more localised to asperity tips with similar levels of deformation seen for both $R_a = 0.25$ μm and $R_a = 0.35$ μm surfaces. Since the stress states for these higher roughness surfaces were similar, the similar resulting plastic zones were expected. This increase in plastic zone at higher surface roughness appeared to be more severe at higher temperature, likely to be due to the reduction in critical resolved shear stress. Increasing the surface roughness reduced the total contacting surface by preventing some asperity contact pairs, increasing the localised load under asperities and therefore plastic zone.

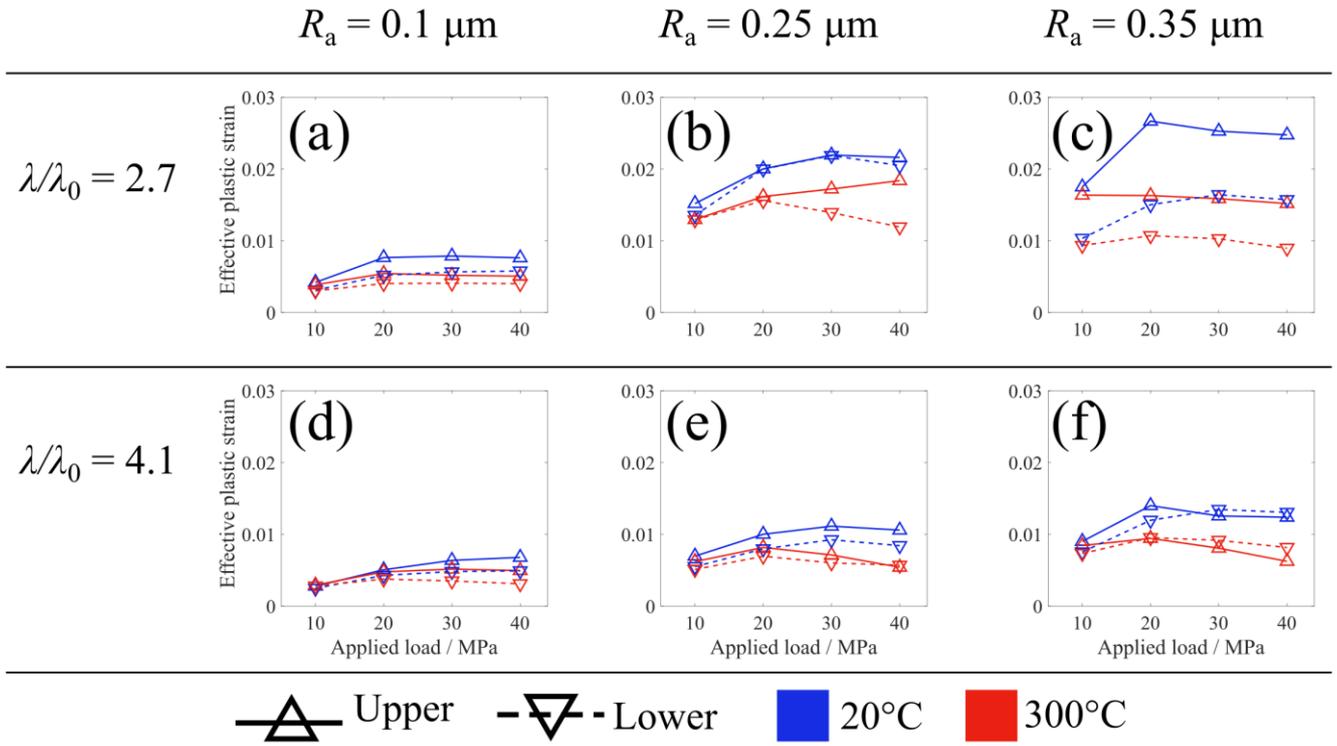

Figure 17: Effective plastic strain for the two temperatures at the end of sliding, averaged over the plastic zone.

Surface roughness caused an increase in plastic strain, as shown in Figure 17. At higher surface roughness values, fewer, sharper asperities supported the same apparent load and therefore deformed to a much greater degree, hence the increase in plastic strain. In the lower roughness surfaces, the contact area was larger, distributing the load over a greater number of asperities and reducing plastic strains. Horizontal scaling also had somewhat of an effect on the effective plastic strain. Again, this was due to the reduction in the number of asperities in contact increasing the localised stress at asperity.

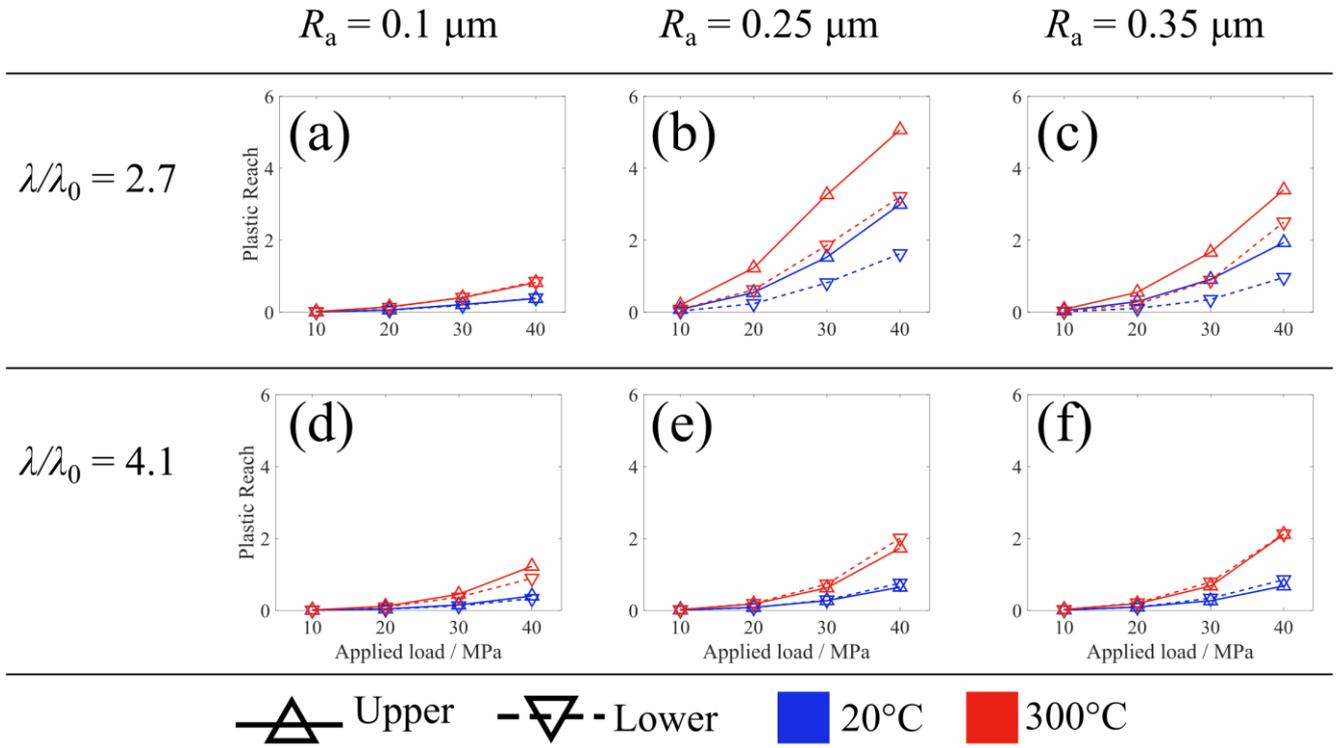

Figure 18: Plastic reach $p_R$ at the end of sliding.

Plastic reach appears to be highly sensitive to the exact surface geometry since both vertical ("surface roughness") and horizontal scaling were found to effect plastic reach values in a non-linear way (Figure 18). Variation in plastic reach due to surface profile far outweighed that due to changes in temperature. This suggests that exact surface geometry is of more importance than temperature in determining the galling resistance of a surface.

### 6.2.3  Galling frequency comparison

Comparisons between the model results and the literature data of Harsha et al. [18] are presented in Figure 19. In general, the experimental trends are captured but quantitative agreement is not achieved. The utility of the results is limited due to the sensitivity of galling to the surface conditions of the tested samples. However, both simulation and experimentation agree in so far as galling resistance decreases with elevated temperature.

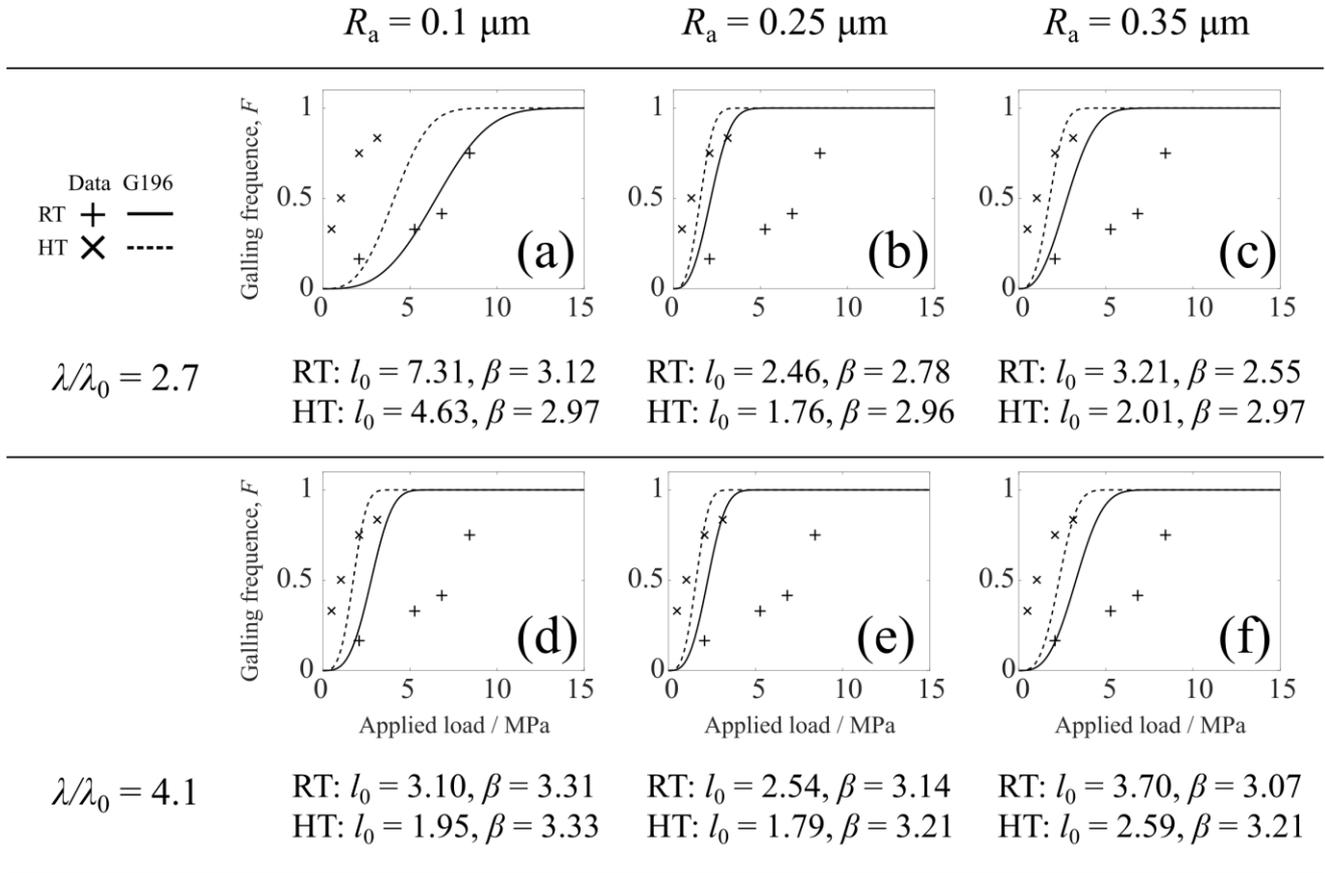

Figure 19: Resulting galling frequency curves, with parameters. All shown with experimental data from [18] with $R_a$ in the range 0.25 – 0.35 μm. Room temperature (RT, solid lines, + ) and 300°C (HT, dashed lines, × ) values for $l_0$ and $\beta$ are given for each surface profile.

For a fixed pair of surfaces, the reduction in characteristic galling load with temperature was much more modest than seen experimentally. Therefore, this suggests that the true mechanism for this temperature sensitivity is not captured by this model. This trend was observed for all surface roughness values.

The nominal surface roughness values investigated here are represented by short sections of a surface profile in 2D (several tens of microns) rather than the full surface topography. Accordingly, the results from the model demonstrate a high level of sensitivity to the exact surface profile since the characteristic galling load is determined from a single asperity contact.

The distribution shape factor $\beta$ was found to be relatively insensitive to both surface and temperature changes. The variation seen in $\beta$ was not large enough to cause changes in the resulting galling frequency curves. For all cases, increases in surface roughness (0.10 → 0.25 μm) resulted in a reduction in characteristic galling load at both ambient and elevated temperatures. However, further increases in surface roughness (0.25 → 0.35 μm) resulted in a subsequent increase in $l_0$. As discussed in Section 6.2.2, the severity of the plastic deformation increased with increasing roughness but saturated above a roughness of 0.25 μm due to the height of asperities preventing contact between portions of the material surfaces.

An interesting difference between surfaces (a) and (d) is the change in $l_0$ value found when the wavelength was scaled. This can be explained by considering the number of asperities in contact during the loading phase. The characteristic galling load was determined by examining the accumulated effective strain in the first asperity pair to make contact. However, subsequent deformation could sometimes bring a second asperity pair into deformation. In (a), the deformation of the first asperity pair during the initial elastic deformation was sufficient to bring a second pair into contact, dividing the load between the two. This reduced the load experienced by the first asperity pair, leading to a higher valve of apparent load required to meet the $p_{\text{eff}} > 0.2$ criterion to ascertain $l_0$. In (d), only one asperity pair made contact during the loading phase, hence taking the full load individually and a lower apparent load was required to cause yielding.

It would be expected that surfaces such as these would display some degree of periodicity. Increasing the length of surface profile considered would bring additional asperities into contact, but the applied normal stress would remain constant and the inclusion of additional repeating units would not provide further insight. One method to determine the separation of asperities is the average profile element length (BS ISO 4287:1997 [58]). These surfaces display an average profile element length of 10 μm. Therefore, these models consider multiple profile elements. However, this does indicate the sensitivity of this method in determining galling resistance to particular surface profiles.

# 7 Conclusions

The thermal effects in galling of internal heat generation and elevated temperature were assessed, along with the role of adhesion strength, using a representative crystal plasticity finite element model, developed from the original work of Barzdajn et al. [31], relating microscopic surface deformation to galling frequency predictions. Contacting asperity deformation was found to be largely insensitive to adhesion strength, with the surface geometry dominating the plastic deformation behaviour. The heating contributions of plastic deformation and friction were examined at representative timescales. The rate of heat transfer far exceeded that of generation, preventing localised heating and resulting in no tangible effects on the deformation of the surfaces and is unlikely to play a role in asperity deformation under these loading conditions.

The effects of isothermal, elevated temperatures were hence investigated in the absence of any deformation related heating. The reduction in the critical resolved shear stress at elevated temperature caused a greater volume of material to undergo plastic deformation in normal and sliding contact; however, a reduction in the average effective plastic strain was observed. The plastic reach showed strong temperature sensitivity. The characteristic galling load showed temperature sensitivity, and this was related to the temperature dependence of the critical resolved shear stress. The Weibull shape factor showed little temperature sensitivity. The resulting galling frequency curves showed reduced galling resistance at elevated temperature, but no abrupt collapse as reported by Kim and Kim [8], suggesting a change in deformation mechanism not captured by this model could be the cause of these experimental observations.

The model results captured the trends of the ASTM G196 tests performed by Harsha et al. [18], when the appropriate surface roughness ($R_a$) values were assessed. The reduction in galling resistance with temperature for a fixed profile was not as large as presented in the literature, again suggesting that some other factor contributes to the reported collapse in galling resistance. This study emphasises that the arithmetic surface roughness alone is a poor measure in quantifying surfaces in contact and galling response.

# 8 Acknowledgements

The authors gratefully acknowledge the support of Rolls-Royce plc., EPSRC and the ICO CDT (EP/L015900/1). DD also acknowledges the support received via his Engineering and Physical Sciences Research Council (EPSRC) Established Career Fellowship (EP/N025954/1) and FPED his Royal Academy of Engineering Research Chair.